# A systematic approach for designing zero-DGD coupled multi-core optical fibers

MIDYA PARTO, MOHAMMAD AMIN EFTEKHAR, MOHAMMAD-ALI MIRI, RODRIGO AMEZCUA-CORREA, GUIFANG LI, DEMETRIOS N. CHRISTODOULIDES*

*CREOL, College of Optics and Photonics, University of Central Florida, Orlando, Florida 32816-2700, USA*
*Corresponding author: demetri@creol.ucf.edu*

**An analytical method is presented for designing N-coupled multi-core fibers with zero differential group delay. This approach effectively reduces the problem to a system of N-1 algebraic equations involving the associated coupling coefficients and propagation constants as obtained from coupled mode theory. Once the parameters of one of the cores are specified, the roots of the resulting N-1 equations can then be used to determine the characteristics of the remaining waveguide elements. Using this technique, a number of pertinent geometrical configurations are investigated in order to minimize intermodal dispersion. © 2016 Optical Society of America**

Since the early days of optical fiber communications, minimizing the differential group delay (DGD) in multimode waveguide structures has been an ongoing issue. This problem was first addressed in 1968 [1] by Kawakami and Nishizawa after recognizing that the parabolic index profile is nearly optimum in terms of substantially reducing intermodal dispersion. What makes this particular type of graded index fiber superior in this regard is the fact that it acts to first order as a perfect achromatic imaging system - thus allowing all possible rays to arrive at the same time [1-4]. More rigorous treatments of this problem based on either modal electromagnetic approaches or WKB methods also lead to similar findings [1,5]. Over the years, a number of techniques have been developed in order to further improve the DGD performance of nearly parabolic multimode optical fibers [6,7]. With the advent of mode division multiplexing [8-13] as a means to meet the rapid-growing needs of the telecommunication industry, there has been a resurgence of interest in designing waveguide structures with reduced modal delay dispersion. This is particularly important in multiple input multiple output (MIMO) signal processing applications where the computational overhead increases linearly with the modal group delay [14].

In recent years multicore fiber structures have emerged as a viable platform for space division multiplexing schemes [15-19]. In this respect, random coupling processes on weakly coupled multicore systems can be exploited to reduce DGD effects [19]. Meanwhile, even though closely-packed arrangements are always advantageous, the resulting proximity of cores eventually proves problematic because the induced cross talk or coupling among waveguide channels can lead to substantial DGD levels between the corresponding supermodes [15, 20-22]. One way to overcome this impediment is to make the coupling coefficient dispersion-free, i.e. frequency-independent. In the case where identical step index fibers are involved, one can show that this is only possible when the operating V number is close to 1.7 provided that the cores are in direct contact with each other [23]. Following this approach, 3-core equilateral multimode fiber designs have been theoretically suggested in the literature with DGDs below 3.2 ps/km. However, as indicated in Ref. [17], the applicability of this tactic is limited, not only in terms of a restricted operating V number, but also because of higher-order (beyond nearest neighbor) cross coupling effects that are prevalent in more complex multicore geometries entailing more elements. Clearly of importance would be to develop alternative methodologies capable of addressing the aforementioned issues in more general coupled multicore fiber (MCF) configurations. Yet, unlike graded index multimode fibers, where an array of analytical and computational tools already exists in order to tackle this problem, finding ways to eliminate DGD in coupled MCFs is still a challenge.

In this paper we show that the resulting eigenvalue polynomial associated with an N-core MCF structure can be appropriately recast into N-1 algebraic equations whose roots can determine the features of the individual waveguide elements needed to achieve zero-DGD conditions. This procedure, based on coupled mode theory, can be employed in a versatile fashion for any MCF arrangement. A number of examples are provided to elucidate this method. By considerably restricting the search space for relevant parameters, these results can be further fine-tuned using finite element methods.

We begin our analysis by considering an N-core MCF system, as shown in Fig. 1 for N = 6. In this respect we assume that the propagation constants of each waveguide involved is $\beta_l$ while the coupling coefficients between different sites is $\kappa_{lm}$. The evolution of the (local) modal field amplitudes $U_l$ in this coupled array is described by the following set of equations:

$$i\frac{dU_l}{dz} + \beta_l U_l + \sum_{l \neq m} \kappa_{lm} U_m = 0, \quad (1)$$

where $l = 1, 2, \ldots, N$. The N eigenvalues $\mu_j$ associated with the supermodes $\bar{U}_j = \bar{U}_{0j} \exp(i\mu_j z)$ of this array can then be directly obtained from the eigenvalue problem,

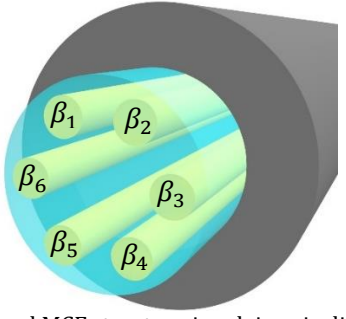

**Fig. 1.** A hexagonal MCF structure involving six dissimilar cores.

$$\begin{vmatrix} \beta_1 - \mu & \kappa_{12} & \cdots & \kappa_{1N} \\ \kappa_{21} & \beta_2 - \mu & \cdots & \kappa_{2N} \\ \vdots & \vdots & \ddots & \vdots \\ \kappa_{N1} & \kappa_{N2} & \cdots & \beta_N - \mu \end{vmatrix} = 0, \quad (2)$$

which in turn leads to the following characteristic equation:

$$\mu^N + P_{N-1}(\beta_l, \kappa_{lm})\mu^{N-1} + P_{N-2}(\beta_l, \kappa_{lm})\mu^{N-2} + \cdots + P_0(\beta_l, \kappa_{lm}) = 0. \quad (3)$$

Here $P_j(\beta_l, \kappa_{lm})$ represent polynomial functions of the propagation constants and coupling coefficients. At this point we emphasize that both parameters $\beta_l$ and $\kappa_{lm}$ are dispersive, i.e. they depend on the optical angular frequency $\omega$. If all the waveguide channels are identical, in which case they exhibit the same propagation constant $\beta_0$, then it is straightforward to show that the resulting eigenvalues are only functions of the inter-site couplings, i.e. $\mu_j = \beta_0 + f_j(\kappa_{lm})$, where $f_j$ represents again a polynomial function. In this particular case DGD can be eliminated only if $d\kappa_{lm}/d\omega = 0$, that is only possible under the conditions mentioned above [17]. Otherwise, a uniform array will suffer from considerable differential group delay levels. Hence, in order to eliminate DGD one has to resort to multicore fiber designs where the constituent waveguide elements are dissimilar. This fact has been recently recognized in connection with zero-DGD two-core structures with unequal radii [24].

Evidently, under zero-DGD conditions, the eigenvalues of the MCF configuration should satisfy the following relation:

$$\frac{d\mu_1}{d\omega} = \frac{d\mu_2}{d\omega} = \cdots = \frac{d\mu_N}{d\omega}, \quad (4)$$

where all derivatives are evaluated at the central frequency $\omega = \omega_0$. Equation (4) ensures that the group velocities of all N supermodes are equal. By differentiating Eq. (3) with respect to $\omega$ we obtain:

$$\mu^{N-1}\left(N\frac{d\mu}{d\omega} + \frac{dP_{N-1}}{d\omega}\right) + \sum_{k=0}^{N-2}\left[(k+1)P_{k+1}\frac{d\mu}{d\omega} + \frac{dP_k}{d\omega}\right]\mu^k = 0. \quad (5)$$

Equation (5) represents a polynomial equation in $\mu$ of degree N-1. For Eq. (5) to be true for all possible eigenvalues $\mu$, it is necessary that the coefficients of all the terms in this polynomial vanish, i.e. $(k+1)P_{k+1}\mu' + P_k' = 0$, where $\mu' = d\mu/d\omega$, etc. By applying this condition to the first term in Eq. (5) and given that $P_{N-1} = -\sum_{l=1}^{N}\beta_l$ we find that:

$$\frac{1}{v_{g_{MCF}}} = \frac{1}{N}\left(\sum_{l=1}^{N}\frac{1}{v_{g_l}}\right), \quad (6)$$

where $v_{g_{MCF}}^{-1} = \mu'$ and $v_{g_l}^{-1} = \beta_l'$. Equation (6) directly implies that the expected common group speed in this MCF system should be in fact the average of all the group velocities associated with the different fiber cores involved. In other words, the common group velocity of all the supermodes in this zero-DGD MCF must be $v_{g_{MCF}}$. From Eq. (5) and after using this latter result, one obtains the following system of N-1 algebraic equations, where $k = 0,1, \ldots, N-2$.

$$\frac{(k+1)P_{k+1}}{N}\left(\sum_{l=1}^{N}\frac{d\beta_l}{d\omega}\right) + \frac{dP_k}{d\omega} = 0. \quad (7)$$

In general, one can introduce the dissimilarity between cores (needed for zero DGD) either by employing different index profiles, by using unequal dimensions, or both. Here, to demonstrate this method, we assume that the core index profiles are all of the same step index type, while we allow the waveguide channel radii to vary around the radius of the first core, which here is held constant. In other words, we use the radii of the N-1 waveguides as variables to solve the N-1 equations of (7). The coupling coefficients between different sites $l, m$ are calculated using the following relation [25]:

$$\kappa_{lm} = (2\Delta_m)^{\frac{1}{2}}\frac{U_l U_m}{R_l V_l} \times \frac{K_0\left(\frac{W_l d_{lm}}{R_l}\right)}{K_1(W_l)K_1(W_m)}$$

$$\times \left\{\frac{\overline{W_l}K_0(W_m)I_1(\overline{W_l}) + W_m K_1(W_m)I_0(\overline{W_l})}{\overline{W_l}^2 + U_m^2}\right\}, \quad (8)$$

where $\overline{W_l} = W_l R_m/R_l$, $I_m(x)$ and $K_m(x)$ are modified Bessel functions of the first and second kind, and $R_l$ are the core radii. Moreover, $\Delta_m = (n_{1,m} - n_{2,m})/n_{1,m}$ is the normalized index difference, $V_l = k_0 R_l n_{1,l}\sqrt{2\Delta_l}$ represents a dimensionless V number for core $l$, while $W_l = R_l(\beta_l^2 - k_0^2 n_{2,l}^2)^{1/2}$, $U_l = R_l(k_0^2 n_{1,l}^2 - \beta_l^2)^{1/2}$, where $n_{1,m}$ and $n_{2,m}$ are the core and cladding refractive indices, $k_0$ is the free-space wavenumber, and $d_{lm}$ is the distance between core centers at sites $l$ and $m$.

Clearly, for a given set of characteristics, both quantities $\beta_1$ and $\beta_1'$ associated with the first waveguide are known. Therefore, the only unknowns in the system of Eqs. (7) are the propagation constants of the remaining N-1 cores and their derivatives with respect to frequency, which are of course functions of the characteristics of the individual waveguides comprising the structure. Here we use the core radii as variables even though in general one can also exploit other degrees of freedom (index contrast, index profile, etc.). It is important to emphasize at this point that the polynomials associated with Eq. (7) directly involve the propagation constants $\beta_l$ (which are inherently large) -hence the resulting eigenvalues $\mu_j$ are of the same order. This in turn complicates the numerical search for roots. To alleviate this problem, we rewrite each propagation constant in terms of a reference value, i.e. $\beta_l = \delta\beta_l + \bar{\beta}$, where in our case $\bar{\beta} = k_0(n_1(\omega) + n_2(\omega))/2$. As a result, the common reference value $\bar{\beta}$ drops out from these equations, and instead, the polynomials $P_j(\delta\beta_l, \kappa_{lm})$ now depend only on the smaller differential propagation constants $\delta\beta_l$. This latter normalization further facilitates the numerical search for solutions. In this normalized frame, the eigenvalues $\mu_j$ are also rescaled with respect to the reference floor. With this in mind, we computationally search for the N-1 radii capable of simultaneously satisfying the N-1 Eqs. (7). In all cases, we seek solutions where each waveguide still remains single-moded, i.e. with a V number less than 2.4. Note that our results, however close to the optimum design, are still not exact due to the approximations inherent in coupled mode theory. Nevertheless, once the search space for relevant core parameters is significantly narrowed down using our approach, the final MCF design can be further fine-tuned using finite element methods.

To demonstrate our method, we first consider a four-core fiber design. The structure is composed of step-index waveguide elements with centers placed on the vertices of a $16\mu m \times 16\mu m$ square (Fig. 2). Here we set the radius of the first core to be $R_1 = 4.42\ \mu m$ and we then determine the remaining cores so that the three equations (of Eq. (7)) are simultaneously satisfied. For this case, the core radii are found to be: $R_2 = 3.929\ \mu m$, $R_3 = 4.095\ \mu m$ and $R_4 = 4.219\ \mu m$. At $\lambda_0 = 1550\ nm$, the resulting DGD between any two of the four supermodes (or eight including polarizations) varies between 0.8 and 7.7 $ps/km$. In other words, zero DGD conditions can only be established provided that all elements are quite dissimilar with respect to each other. The mode intensity profiles of the corresponding supermodes are depicted in Fig. 3.

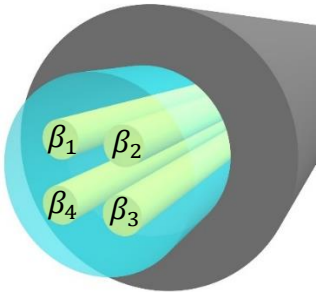

**Fig. 2.** Schematic of a four-core MCF.

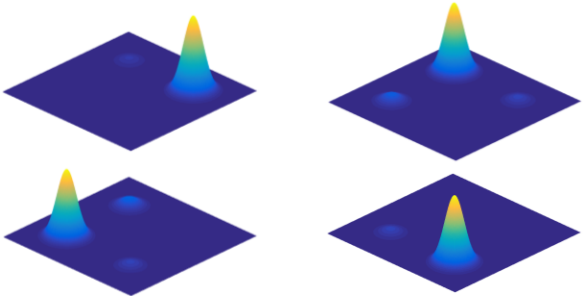

**Fig. 3.** The four supermode intensity profiles of an ultra-low DGD four-core MCF design.

For the scenario explored above, Fig. 3 clearly indicates that each supermode in this four-element system mostly resides in one core-with little penetration into the other three. This characteristic could be desirable in terms of multiplexing or demultiplexing these modes, in particular in connection to MIMO processing. The effective mode indices corresponding to the four modes shown in Fig. 3 are 1.4463, 1.4462, 1.4461 and 1.4459, while the core and cladding refractive indices are taken here to be 1.449 and 1.444 respectively. Our results suggest that zero DGD is possible even though the different supermode propagation constants are clustered away from the cladding radiation modes of the structure. To assess the efficacy of our method, we compare the DGD values we found to those associated with a similar four-core arrangement, comprised of identical waveguides with a radius of $R_1 = 4.42\ \mu m$ (all other characteristics remain the same). For this uniform array, the maximum DGD was 2.3 ns/km, which is nearly 300 times higher than the values obtained for the aforementioned design.

The method presented here is general and can be applied to MCFs with different numbers of cores and geometries. The results for MCFs with two, three, five and six cores, of different geometries, are presented in Table 1. In these examples, all the parameters of the MCF structure (e.g. core and cladding refractive index, center frequency, spacing between elements, etc.) are the same as for the four-core case studied above, and the radii of N-1 cores are chosen as unknowns in simultaneously solving the N-1 equations of (7). The table provides the V numbers of the individual cores in isolation. The maximum expected value of DGD between the various supermodes of the structure, as well as the mode intensity profiles for each supermode is also presented. As in the four-core example considered before, even in more complex geometric patterns, all the core elements must be different when taken in sequence, to achieve low DGD conditions. Moreover, each supermode tends to primarily occupy one waveguide element. Other MCF systems can be analyzed in a similar fashion based on the analytical scheme suggested. Even though in our examples we considered step-index elements so as to demonstrate this approach, in principle one can apply the same methodology to more involved graded index cores provided the respective propagation constants and coupling coefficients are evaluated as a function of wavelength.

Moreover, we have considered how the DGD varies as a function of wavelength for the aforementioned designs. As an example, we have numerically obtained the three DGD curves associated with the equilateral three-core design in Table 1. This response was determined across the C-band, around the operating wavelength of $\lambda_0 = 1550\ nm$ (Fig. 4). Our computations indicate that throughout the C-band, the maximum DGD remains always below 28.4 $ps/km$. Compared to other equilateral equal-core step-index designs of a similar index contrast, our results show a ten-fold improvement in DGD variations. We note that for this particular design the DGD curves happen to be polarization insensitive. Similar conclusions can be drawn for the other designs in Table 1.

**Table 1. Low DGD MCF structures of different geometries**

| Configuration | Maximum DGD | Waveguide V numbers |
|---|---|---|
| 1—2 (two cores, linear) | 0.5 ps/km | $V_1 = 2.20$ $V_2 = 2.06$ |
| triangle (1, 2, 3) | 0.6 ps/km | $V_1 = 2.20$ $V_2 = 2.37$ $V_3 = 2.10$ |
| 2—1—3 (linear three cores) | 0.5 ps/km | $V_1 = 2.20$ $V_2 = 2.32$ $V_3 = 2.10$ |

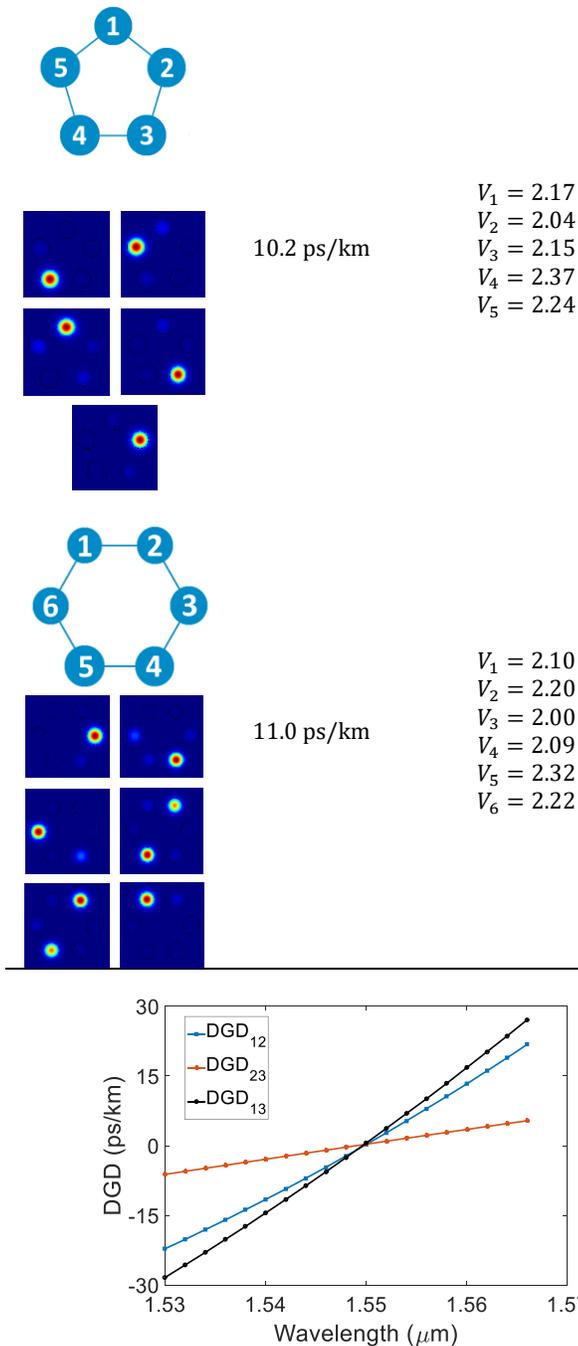

**Fig. 4.** Variation of DGD over the C-band for a three-core fiber design.

In summary, we have developed a systematic approach for designing zero-DGD MCFs. The technique makes use of coupled mode theory to describe the frequency dependent coupling among the elements of the array, and provides the necessary conditions for achieving equal group velocities for the different supermodes. These conditions are expressed in the form of a set of algebraic equations entailing the propagation constants of the individual guides and their coupling coefficients. Once one of the cores is specified, the rest can be determined by solving the aforementioned system of algebraic equations. We applied our method to a variety of MCF configurations. In all of the cases, the resulting DGD levels were exceedingly low.

**Acknowledgment**
This work was partially supported by the Office of Naval Research (MURI Grant No. N000141310649).